\documentclass[conference]{IEEEtran}
\IEEEoverridecommandlockouts
\makeatletter
\def\ps@headings{%
\def\@oddhead{\mbox{}\scriptsize\rightmark \hfil \thepage}%
\def\@evenhead{\scriptsize\thepage \hfil \leftmark\mbox{}}%
\def\@oddfoot{}%
\def\@evenfoot{}}
\makeatother
\usepackage{amsthm}
\makeatletter
\def\th@plain{%
\thm@notefont{}% same as heading font
\itshape % body font
}
\def\th@definition{%
\thm@notefont{}% same as heading font
\normalfont % body font
}
\makeatother
\usepackage{amsfonts}
\usepackage{cite}
\usepackage{mathrsfs}
\usepackage{graphicx}
\usepackage{amsmath}
\usepackage{times}
\usepackage{bm}
\usepackage{multirow}
\usepackage{latexsym}
\usepackage{amssymb}
\usepackage[center]{caption2}
\usepackage{stfloats}
\usepackage{cases}
\usepackage{array}
\usepackage{setspace}
\usepackage{fancyhdr}
\usepackage{graphicx}
\usepackage{subfigure}
\usepackage{balance}
\usepackage{makecell}
\usepackage[ruled,vlined]{algorithm2e}

\ifCLASSINFOpdf
\else
\fi
\begin{document}
\bibliographystyle{IEEEtran}
\title{A Novel High-Rate Polar-Staircase Coding Scheme}
\author{\IEEEauthorblockN{Bowen~Feng, Jian~Jiao, Liu~Zhou, Shaohua~Wu, Bin Cao, and Qinyu~Zhang\\
\IEEEauthorblockA{
Communication~Engineering~Research~Center, Harbin~Institute~of~Technology~(Shenzhen), Shenzhen, China\\
\{hitfbw@hotmail.com; zhouliu@stu.hit.edu.cn; \{jiaojian, hitwush, caobin, zqy\}@hit.edu.cn\}}
\thanks{This work was supported in part by the National Natural Sciences Foundation of China (NSFC) under Grant 61771158, Grant 61525103, and Grant 61371102, and in part by the Shenzhen Basic Research Program under Grant JCYJ20170811154309920, Grant JCYJ20170811160142808, Grant ZDSYS20170728090330586, and Grant JCYJ20160328163327348.}
}}
% make the title area
\maketitle
\IEEEpeerreviewmaketitle
\begin{abstract}
The long-haul communication systems can offer ultra high-speed data transfer rates but suffer from burst errors. The high-rate and high-performance staircase codes provide an efficient way for long-haul transmission. The staircase coding scheme is a concatenation structure, which provides the opportunity to improve the performance of high-rate polar codes. At the same time, the polar codes make the staircase structure more reliable. Thus, a high-rate polar-staircase coding scheme is proposed, where the systematic polar codes are applied as the component codes. The soft cancellation decoding of the systematic polar codes is proposed as a basic ingredient. The encoding of the polar-staircase codes is designed with the help of density evolution, where the unreliable parts of the polar codes are enhanced. The corresponding decoding is proposed with low complexity, and is also optimized for burst error channels. With the well designed encoding and decoding algorithms, the polar-staircase codes perform well on both AWGN channels and burst error channels.
\end{abstract}
\begin{IEEEkeywords}
Polar code, staircase code, concatenation, long-haul communication.
\end{IEEEkeywords}
%\begin{keywords}
%Polar code, space-time block coding, fading channel, construction algorithm
%\end{keywords}

\section{Introduction}
The long-haul communication systems, such as the Ka-band high throughput satellites and free-space optical communications, can offer ultra high-speed data transfer rates by using millimeter wave even optical wave.
However, such extreme short waves often suffer from burst errors due to the signal loss by air turbulence, clouds or precipitation and solar wind \cite{hasegawa2017transmission,kyrgiazos2014gateway}.
The burst error channels can be modeled as a two-state Markov model called Gilbert-Elliott model, which is defined in \cite{mushkin1989capacity}.
In order to transmit reliable on burst error channels, some techniques have been applied. The most common technique is the hybrid automatic repeat request (HARQ) scheme \cite{ghosh2005broadband}.
However, for long-haul transmission, the large latency of the retransmission make the HARQ scheme do not work well.

Staircase code is a new technique which has the potential to improve the reliability of the long-haul transmission \cite{liga2017information}.
It is a product-like code, constructed in a staircase form as its name suggests.
It has been tested for high-speed optical communications owing to the high-rate and high-performance features \cite{smith2012staircase}.
We can regard the staircase coding scheme as a concatenation structure. The component forward error correction (FEC) codes which build the staircase structure are concatenated by the overlapping parts.
The LDPC-staircase codes have been researched extensively in \cite{zhang2014staircase,zhang2017complexity}. However, they suffer from the high decoding complexity due to the high-complexity internal belief propagation (BP) decoder for the component LDPC codes and the high-complexity external iterative sliding window decoder for staircase codes.

Polar coding, introduced by Ar\i kan in \cite{arikan2009channel}, has been proved to achieve the capacity of the memory-less symmetric channel with low-complexity encoder and decoder. The high-reliability polar codes are considered to be applied for eMBB services in the 5G communication system. However, the reliability of polar codes could not be maintained as the code rate rises, because that more incompletely polarized parts are applied to contain information bits. In order to improve the performance of high-rate polar codes, the concatenation of polar codes and other channel codes have aroused many researchers' interests. In fact, the outstanding cyclic redundancy check-aided successive-cancellation list (CRC-aided SCL) decoding \cite{tal2011list} is a concatenation of polar codes and CRC codes, which provides superior bit/block error rate (BER/BLER) performance.

The polar codes have been also concatenated with BCH codes \cite{wang2014concatenations}, Reed-Solomon (RS) codes \cite{mahdavifar2014performance}, convolutional codes and LDPC codes \cite{zhang2014polar}. We have proposed a concatenation scheme of polar codes and space-time block codes for MIMO systems in \cite{feng2016construction}. All of the above concatenations can outperform than independent polar codes. However, some of them suffer high encoding and decoding complexity, and the overall rate is unsatisfactory due to the multi-layer concatenations. Thus, a new idea enters our mind to concatenate the polar codes with themselves. The staircase coding make it possible. In staircase decoding, the soft decoding of the component codes is required. The BP and the soft cancellation (SCAN) are the two impactful soft decision decoding algorithms for polar codes \cite{fayyaz2014low}. The SCAN decoding can achieve the same performance with BP decoding with less iterations, and the complexity of the SCAN decoding is far less than the one of the BP decoding. Thus, it is meaningful to do the attempt to study the polar-staircase codes. Not only can it reach the high-rate and high-performance, but the complexity of it can be low.
%SCAN比BP复杂度低，性能差不多，那就是比SCL的性能差？
%Thus, a new idea enters our mind that is: Why not concatenate the polar codes with themselves?

%We have made a preliminary attempt to concatenated the polar codes in [], where we propose a rateless polar coding scheme to achieve arbitrary code rate. The scheme provides a rational rate adapting to the time-varying characters of channels, but it does little to improve the reliability of the codes. Thus, it is not a whole concatenation for polar codes.

Our aim in this paper is to propose a novel polar-staircase coding scheme as an effective way to be appropriate for high-speed long-haul communications. We will adopt the systematic polar codes as component codes to build the staircase form, which will help to separate the information bits and the check bits. The check bits and the unreliable parts of the information bits will compose the overlapping parts in order to improve the BER/BLER performance of the polar-staircase codes. The main contributions of this paper are stated as follows. Firstly, a polar-staircase coding scheme is proposed. The well designed encoding algorithm of the polar-staircase codes will be described in detail. Secondly, the decoding algorithm of the polar-staircase is proposed, where the condition of burst error channel is also considered. The complexity is analyzed and compared with the state-of-the-art LDPC-staircase scheme. The SCAN decoding algorithm for systematic polar codes is also described as a basic ingredient. Finally, the performance simulations of polar-staircase codes are provided on both AWGN channels and burst error channels. The comparisons with the state-of-the-art staircase codes will be also considered.

The remainder of the paper is organized as follows. Section \ref{Preliminaries} describes the preliminaries of systematic polar codes and staircase codes. Section \ref{analysis} proposes the encoding and decoding of the polar-staircase coding scheme. The complexity analyses is also described. Simulation results of the polar-staircase coding scheme are provided in Section \ref{simulations}. Finally, Section \ref{conclusion} concludes the paper.

\section{Preliminaries}\label{Preliminaries}
\subsection{Density Evolution for Polar Codes}
The process of channel polarization is provided in \cite{arikan2009channel}. The polarized channel $W_N$ combined by $N$ channels $W$ is split to parallel subchannels $\{W_N^{(i)}\}$ whose capacities are different.
Density evolution (DE) is a common way to evaluate the reliability of each subchannel/coding position \cite{mori2009performance}. The high-reliability positions are select to carry the information bits, where their indices form an information set $A$. The rest carries the frozen bits, and their indices form an frozen set $A^c$. The selection of information set is a key factor that impacts the performance of polar codes.
The DE guides the selection of information set by ranking the probabilities of incorrect messages of subchannels.
The probabilities of incorrect messages can be obtained by calculating the probability density functions (pdfs) of log-likelihood ratios (LLRs) passing in the coding graph, where the pdfs are regarded as the densities. We can treat the LLR of the $i$-th subchannel $W_N^{(i)}$ as a variable, and the pdf of the variable is expressed as $a_N^{(i)}(z)$. We assumed that all zero bits are transmitted and the channel $W$ is symmetric, and then the probability of incorrect messages of the $i$-th subchannel can be expressed as $P_e(i)=\int_{-\infty}^{0}a_N^{(i)}(z)\textrm{d}z$. The densities passing in the successive-cancellation (SC) decoding graph can be calculated as follows,
\begin{equation}
a_{2N}^{(2i)}=a_{N}^{(i)}\star a_{N}^{(i)},~~a_{2N}^{(2i-1)}=a_{N}^{(i)}\boxdot a_{N}^{(i)},~~a_{1}^{(1)}=a_{W}
\label{densities}
\end{equation}
where $a_W$ is the pdf of the initial channel $W$'s LLR when 0 is transmitted, and where $\star$ and $\boxdot$ are the convolution operations for variable nodes and check nodes respectively \cite{richardson2008modern}. When densities of all subchannels are obtained, the corresponding probabilities of incorrect messages can be acquired by calculation. Gaussian approximation (GA) is often adopted as a substitution for the DE. The computational complexity of GA is lower than the one of the DE.

\subsection{SCAN Decoding of Systematic Polar Codes}
The SCAN decoding is an efficient soft decoding of polar codes. It can be regarded as the combination of the SC decoding and the BP decoding. In each iteration, the LLR information $L$ is updated from the channel observations to the source bits, then the LLR information $R$ is updated reversely. The recursion formulas adopted in updating the LLR information are the same with the ones in BP decoding, but the update order of the LLR information follows the example of the SC decoding. We can describe the SCAN decoding as a decoding with the \textit{branch} of SC and the \textit{leaves} of BP. In \cite{fayyaz2014low}, the algorithm of the SCAN decoding is provided. In practice, we recommend the BP recursion formulas provided in \cite{yuan2013architecture}, which are easier for implementing.

The systematic polar coding can be treated as a two-step polar coding as shown in \cite{sarkis2016flexible}. The source bits are coded twice continuously with assigning zeros to all the frozen bits of every time. The information bits of the systematic polar codeword are the same with the inputting bits. The check bits of the systematic polar codeword can be expressed as
\begin{equation}
x_{A^c}=uG_{AA}G_{AA^c},
\label{sysfroz}
\end{equation}
where the generator matrix $G$ is obtained by the recursion of the Kronecker products of the kernel matrix without bit-reversing, and where $G_{AA^c}$ denote the matrix consist of the elements in the $i,i\in A$ column and $j,j\in {A^c}$ row of $G$. The conventional decoding of polar codes can also be applied for systematic polar codes. The source bits can be recovered by multiplying the output of the decoder with the generator matrix. In this work, the soft information of the component codes should be also applied in the decoding of the staircase form. Thus, we adjust the main algorithm to output the soft information. The SCAN decoder of systematic polar codes is shown in Algorithm~\ref{alg:1}.
\begin{algorithm}
	\caption{SCAN decoder for systematic Polar codes}
	\label{alg:1}
    \KwIn{input $(n+1)\times N$ matrices $L1,R1,L2,R2$, maximum iteration $I_{iter}$, sets $A$ and $A^c$, LLRs $L$ from channel}
%	\begin{algorithmic}[1]
		%\REQUIRE Code length $N_k$, code rate $R_k$, prior information set $A_{k-1}$
%		\ENSURE Information set $A_k$
         $\{L1(1,i),i\in(1,\ldots,N)\}\leftarrow L$\\
         $\{R1(n+1,i),i\in A^c\}\leftarrow \infty$, $\{R1(n+1,i),i\in A\}\leftarrow 0$\\
        \For{$i=1\rightarrow I_{iter}$}
        {
        \For{$\phi=1\rightarrow N$}{$L1\leftarrow$ updatellrmap$(L1)$\\
            \If{$\phi$ is even}{$R1\leftarrow$ updatebitmap$(R1)$}
        }
        }
        $\{L2(1,i)\}\leftarrow \{L1(n+1,i)\}, i\in(1,\ldots,N)$\\
        $\{R2(n+1,i),i\in A^c\}\leftarrow \infty$, $\{R2(n+1,i),i\in A\}\leftarrow 0$\\
         \For{$i=1\rightarrow I_{iter}$}
        {
        \For{$\phi=1\rightarrow N$}
        {$L2\leftarrow$ updatellrmap$(L2)$\\
            \If{$\phi$ is even}
            {
                $R2\leftarrow$ updatebitmap$(R2)$\\
            }
        }
        }
\KwOut{output $L1,R1,L2,R2$}
\end{algorithm}

%\begin{equation}
%\begin{split}
%&R_{\lambda-1}(\psi,2\omega)=sign(R_{\lambda}(\phi-1,\omega))sign([R_{\lambda}(\phi,\omega)+L_{\lambda-1}(\psi,2\omega+1)])\\
%&R_{\lambda-1}(\psi,2\omega)=sign(R_{\lambda}(\phi-1,\omega))sign([R_{\lambda}(\phi,\omega)+L_{\lambda-1}(\psi,2\omega+1)])\\
%\end{split}
%\end{equation}

\subsection{Staircase Codes}
\begin{figure}[t]
\centering
\includegraphics[width=0.38\textwidth]{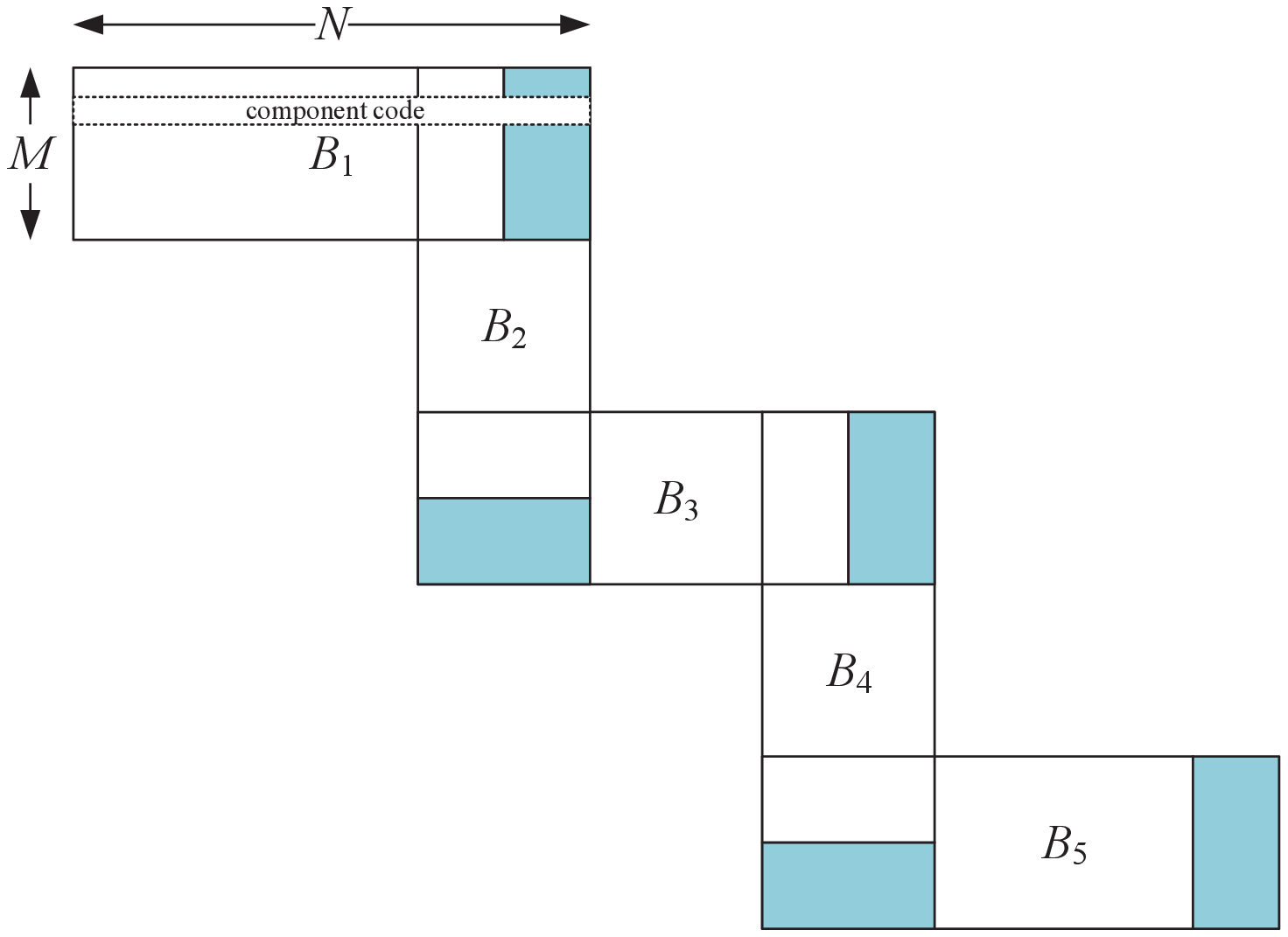}
\caption {Staircase code structure.} \label{f:fig1}
\end{figure}

The staircase codes are constructed by component codes as shown in Fig.~\ref{f:fig1}. Each stair $B_i$ is constituted by $M$ parallel component code-blocks with the same length $N$. The component code-blocks are the same kind of systematic codes. The overlapping part in the stairs $B_i$ and $B_{i+1}$ is transmitted twice. The check bits and a part of the information bits of the stair $B_i$ will be transmitted as information bits in the stair $B_{i+1}$. In order to unify the performance of each stair, the left $M$ columns of the first stair are often set as zeros. The width $M$ of a stair is usually set as $M=N/2$ in previous works to ensure that every bit will be transmitted twice. In this work, we consider that the width $M$ can be set in a range as $M\in(N(1-R),N/2]$, which improve the whole rate of the staircase codes and meets the need of correcting burst errors.

The decoding of the staircase codes is an iterative sliding window structure. At each iteration, the window slides from the last stair to the first stair. When the sliding window reach the stair $B_i$, it decodes each row of the stair by using the component code decoder, then pass the decoding information to the next stair. Thus, the component codes with soft decoders are more suitable for constructing the staircase codes.

\section{Polar-Staircase Coding Scheme}\label{analysis}
\subsection{Construction of Polar-Staircase Codes}
\begin{figure}[htbp]
\centering
\includegraphics[width=0.4\textwidth]{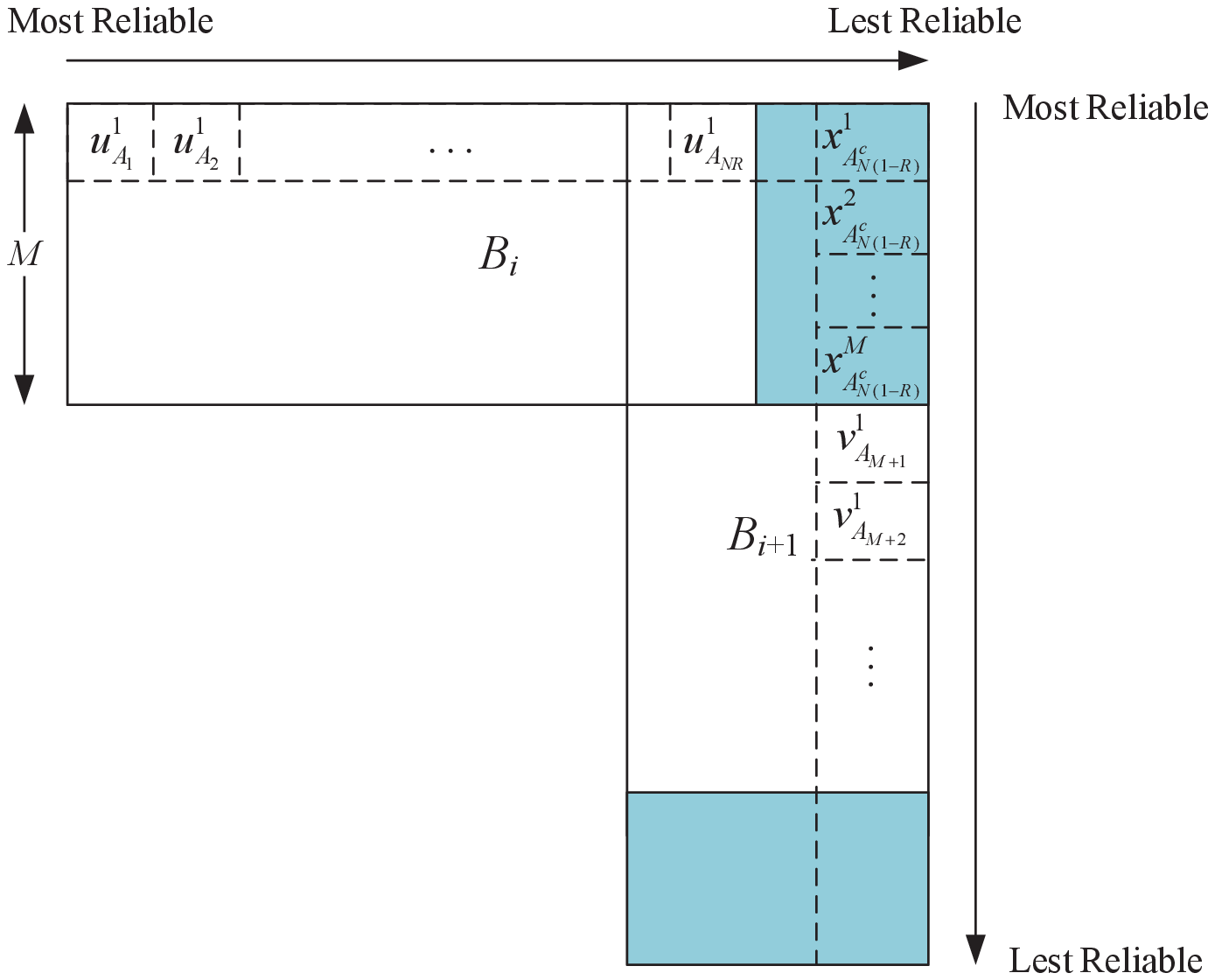}
\caption {Polar-staircase construction.} \label{f:fig2}
\end{figure}

In the polar-staircase coding scheme, we adopt the DE to evaluate the reliability of each coding position, then sort the indices according to the reliabilities. We select the information set $A$ and frozen set $A^c$ by following the descending order of the corresponding reliabilities.
Thus, we consider to enhance the performance of the unreliable bits when designing the concatenation scheme. We show an example of two concatenated stairs $B_i$ and $B_{i+1}$ in Fig.~\ref{f:fig2}.
The stair $B_i$ consists of $M$ systematic polar code-blocks. When the blocks are generated, the rows of the stair should be rearranged according to the reliability of each coding position.
The rows containing the information bits need to be rearranged as $(u_{A_1}^j,u_{A_2}^j,\ldots,u_{A_{NR}}^j)$, and the rest rows containing the check bits need to be rearranged as $(x_{A_1^c}^j,x_{A_2^c}^j,\ldots,x_{A_{NR}^c}^j)$.
The overlapping part $(u_{A_{N-M+1}}^j,\ldots,u_{A_{NR}}^j,x_{A_1^c}^j,\ldots,x_{A_{NR}^c}^j)^\mathrm{T}$ is placed at the $M$ most reliable rows of the stair $B_{i+1}$. With $(NR-M)$ rows of new information bits, the $M$ systematic codes in stair $B_{i+1}$ can be generated. The polar-staircase codes can be constructed with repeating the operation steps above.

 %lest reliable-->least

\subsection{Decoding of Polar-Staircase Codes}
We have provided the SCAN decoding of systematic polar codes in Section~\ref{Preliminaries}. The soft information is available after decoding the component codes. Then the soft information can be applied in the subsequent decoding. For example, in a $5$-stair polar-staircase codes profiled in Fig.~\ref{f:fig1}, the $M$ component systematic polar code-blocks of the last stair $B_5$ should be decoded by the one iteration SCAN decoder. Then the obtained decoding information of the overlapping parts of $B_5$ and $B_4$ need to be combined with the corresponding channel observations of the stair $B_4$. In this work, we adopt a naive way to add the decoding information with the corresponding channel observations of the former stair. After updating the channel observation in stair $B_4$, the $M$ component codes continue to be decoded by the SCAN decoder. The operation steps can go forward until the first stair $B_1$ being decoded. Now, one iteration of the staircase codes is completed. In the subsequent iterations, the soft information of all the stairs is updated periodic. When reaching the maximum iteration $I_{\max}$, the decisions of all the sources can be calculated based on the final soft information. The detailed decoding steps are shown in Algorithm~\ref{alg:2}.
\begin{algorithm}
	\caption{Polar-staircase decoder}
	\label{alg:2}
 %   $\{L_\eta((1:M),(1:N))\} \leftarrow B_{\eta}$\\
%	\begin{algorithmic}[1]
		%\REQUIRE Code length $N_k$, code rate $R_k$, prior information set $A_{k-1}$
%		\ENSURE Information set $A_k$
        \For{$i=1\rightarrow I_{max}$}
        {
        $L_\eta \leftarrow$ LLRs from each stair $B_{\eta}$\\
        \For{$\eta=k\rightarrow 1$}
        {
         $L_{\eta}((1\!:\!M),(N\!\!-\!\!M\!\!+\!\!1\!:\!N))\!=\! R2_{\eta+1}^{M:1}(n\!\!+\!\!1,1\!:\!M)+L_{\eta}((1\!:\!M),(N\!\!-\!\!M\!\!+\!\!1\!:\!N))$

            \For{$m=1\rightarrow M$}
            {
                $\{L1_\eta^m,R1_\eta^m,L2_\eta^m,R2_\eta^m\}=\text{SCANdecoder\_sys}(L1_\eta^m,R1_\eta^m,L2_\eta^m,R2_\eta^m,$\\
                ${L_\eta(m,:),I_{iter}=1)}$
            }
        }
        }
         \For{$\eta=k\rightarrow 1$}
        {
            \For{$m=1\rightarrow M$}
            {
            decision$(L2_\eta^{m}(n+1,:))$
            }
        }
\end{algorithm}

We do not consider the channel with burst errors in the above decoding of polar-staircase codes. When applied in the long-haul transmissions, the decoding should have the ability to reduce the burst errors.
The polar-staircase codes have the ability to fix parts of the burst errors, because the overlapping parts of the codes are interleaved automatically. When the receiver detects that burst errors happen to the overlapping parts, it can adopt the corresponding channel observations form the previous (or next) stair instead of the errors. Thus, when the burst error channel is considered, the extra steps are inserted prior to the polar-staircase decoding as shown in Algorithm~\ref{alg:3}.

\begin{algorithm}
\caption{Polar-staircase decoder for burst error channel}
	\label{alg:3}
    \For {$\phi=1\rightarrow k$}
    {
        $(\tilde{m}_{\phi},\tilde{n}_{\phi}) \leftarrow$ indices of the burst errors detected in $B_\phi$\\
        \If {$\tilde{n}_{\phi}\leq M$}
        {
            $B_\phi(\tilde{m}_{\phi},\tilde{n}_{\phi})\leftarrow B_{\phi-1}(\tilde{n}_{\phi},N-M+\tilde{m}_{\phi})$\\
        }
        \ElseIf {$\tilde{n}_{\phi}> N-M$}
        {
            $B_\phi(\tilde{m}_{\phi},\tilde{n}_{\phi})\leftarrow B_{\phi+1}(\tilde{n}_{\phi}-N+M,\tilde{m}_{\phi})$\\
        }
    }
    Polar-staircase decoder($B$)\\
\end{algorithm}

\subsection{Complexity  Analyses of Polar-Staircase Scheme}
\begin{table}[b]
\centering
\setlength{\abovecaptionskip}{0pt}
\setlength{\belowcaptionskip}{10pt}
\caption{Complexity Comparison of the Decoding of Polar-Staircase and LDPC-Staircase}
\label{table1}
\begin{tabular}{c
!{\vrule width1.2pt}c|c
}
\Xhline{1.2pt}
\multirowthead{3}&
\multicolumn{2}{c}{\thead{Complexity}}\\
\Xhline{1.2pt}
\thead{Decoding \\ Operations} & \thead{Polar-staircase} & \thead{LDPC-staircase}\\
\Xhline{1.2pt}
\thead{Sign/Comparison} & \thead{$6kMN\log N$} & \thead{$kMN(1-R)(d_c+1)$}\\
\Xhline{0.8pt}
\thead{Multiplications} & \thead{$2kMN\log N$} & \thead{$kMN(1-R)(d_c-1)$}\\
\Xhline{0.8pt}
\thead{Divisions} & \thead{$0$} & \thead{$kMN(1-R)d_c$}\\
\Xhline{0.8pt}
\thead{Additions} & \thead{$2kMN\log N+kM^2$} & \thead{$2kMNd_v+kM^2$}\\
\Xhline{1.2pt}
\thead{Total} & \thead{$10kMN\log N+kM^2$} & \thead{$5kMNd_v+kM^2$}\\
\Xhline{1.2pt}
\end{tabular}
\end{table}

The complexity of the SCAN decoding of polar codes is provided in \cite{fayyaz2014low}. We would also like to give a list to analyze the complexity of the polar-staircase scheme more detail and make a comparison with the LDPC-staircase scheme. As shown in Table~\ref{table1}, the complexity comparison of the $k$-stair schemes is provided. We consider in the comparison that the BP decoding of LDPC codes and the SCAN decoding of polar codes are both with one iteration, and the iteration of the sliding window decoding is one, too. According to \cite{fayyaz2014low} and our experiments, the performance of the LDPC codes with $30\sim60$ BP iterations is similar with the one of the polar codes with $4\sim8$ SCAN iterations when the rate is 0.5. For example, we can make a comparison between the polar codes with length $N_p=2048$ and the WiMax standard LDPC codes with length $N_l=2016$ and $d_v=3.33$. The complexity of polar-staircase in one iteration is about six times higher than the one of LDPC-staircase, however the corresponding iterations the polar scheme needed is more than six times lower than the iterations the LDPC scheme needed. The complexity of LDPC codes is higher than the one of polar codes to achieve the similar performance, though the complexity of LDPC codes in one iteration is lower. We will also provide the performance comparisons of LDPC-staircase and polar-staircase in the next section to verify it.

%\subsection{Performance Analyses of Polar-Staircase Codes}

\section{Simulation Results}\label{simulations}
\begin{figure}[b]
\centering
\includegraphics[width=0.45\textwidth]{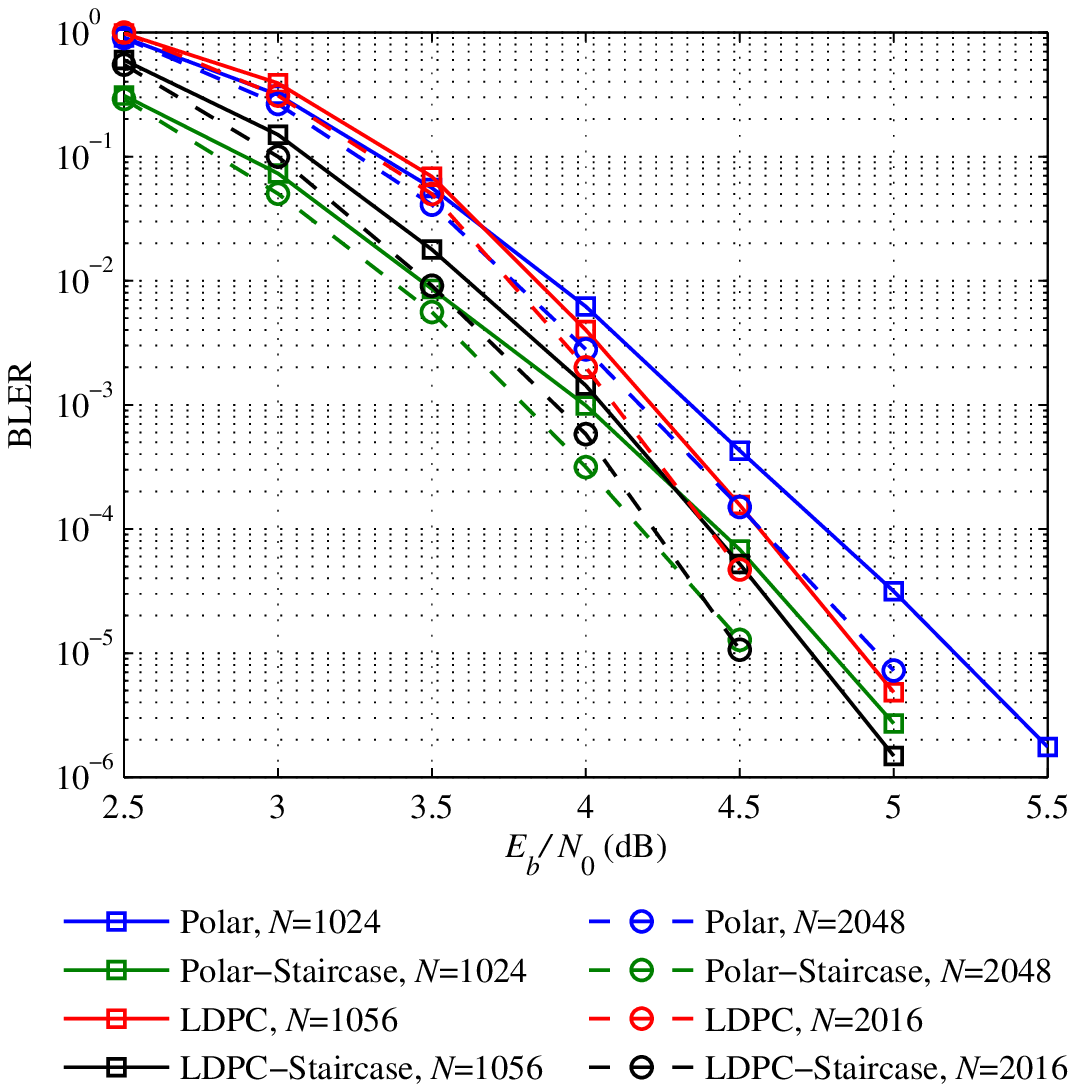}
\caption {BLER performance of polar-staircase codes and LDPC-staircase codes over AWGN channels.} \label{f:fig3}
\end{figure}

In this section, we mainly focus on the BLER performance of the polar-staircase scheme. Firstly, we consider to simulate the performance of the polar-staircase codes in binary input AWGN channels, and make comparisons with the LDPC-staircase scheme. In the polar-staircase scheme simulations, the length of the component systematic polar codes is as $N_p=1024, 2048$ with the code rate $R=5/6$. The corresponding component LDPC codes' length is as $N_l=1024, 2016$ with the same rate $R=5/6$ as described in the WiMax standard. We assume that the width of each stair is as $M=300$ when $N_p=1024$ and $N_l=1056$, $M=600$ when $N_p=2048$ and $N_l=2016$. In order to balance the complexity of the two schemes approximately, the iteration of the polar-staircase decoding is $4$ and the one of the LDPC-staircase decoding is $25$. As shown in Fig.~\ref{f:fig3}, the rate $R=5/6$ polar codes can not perform better than LDPC codes when the $E_b/N_0$ is higher than 3.5dB. However, the polar-staircase concatenation scheme gains more improvement on the BLER performance than LDPC-staircase scheme. The polar-staircase codes can outperform than LDPC-staircase codes at low $E_b/N_0$, and the polar-staircase scheme with $N_p=2048$ can reach the BLER $10^{-4}$ sooner than the LDPC-staircase scheme. It is because that the unreliable parts of the polar codes measured by DE are mostly concatenated, and they are enhanced by the reliable parts of the next stair.

\begin{figure}[h]
\centering
\includegraphics[width=0.44\textwidth]{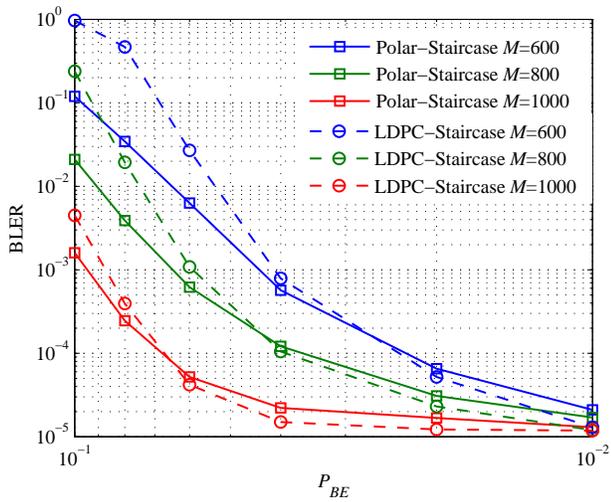}
\caption {BLER performance of polar-staircase codes and LDPC-staircase codes over burst error channels.} \label{f:fig4}
\end{figure}

Moreover, we will consider the condition that the polar-staircase codes are transmitted over the $E_b/N_0=5$dB AWGN channel with burst errors. The Gilbert-Elliott model is adopted in the simulations with the average burst error length $\triangle=20$ and the average error probability $P_{BE}$. The LDPC-staircase is still adopted in the simulations as comparisons. We assume that the length of the component polar codes is as $N_p=2048$, and the rate is as $R=5/6$. The corresponding component LDPC codes are with the length $N_l=2016$. In the following three group of simulations, the width of each stair is assumed as $M=600, 800, 1000$, respectively. As shown in Fig.~\ref{f:fig4}, the polar-staircase codes can provide better BLER performance than LPDC-staircase codes when $M=600$ and $P_{BE}\geq0.04$. However, the advantage decreases as the width $M$ rises. The LDPC-staircase codes perform better when $M=1000$ and $P_{BE}\leq0.06$. We can find that the polar-staircase codes can provide better performance than LDPC-staircase codes at high error probabilities, but the performance of the LDPC-staircase codes improves faster as the number of the errors decreases.

\section{Conclusions}\label{conclusion}
In this paper, we propose a novel high-rate polar-staircase coding scheme, which has the potential to be applied for long-haul communications to achieve high-speed and high-performance. We provide an elaborate encoding algorithm of the polar-staircase codes, which enhance the unreliable parts of the polar codes. The low-complexity soft decoding algorithm of the polar-staircase codes is proposed with combining the SCAN decoding and iterative sliding window decoding. The high-rate polar-staircase codes can perform well over AWGN channels. They can outperform than LDPC-staircase codes at low $E_b/N_0$. The polar-staircase codes can also provide better performance than LDPC-staircase codes at high error probabilities, but the performance of the LDPC-staircase codes improves faster as the number of the errors decreases. In subsequent work, we will test the polar-staircase coding scheme in a real long-haul scenario and optimize the decoding in order to improve the performance further.

%\section*{Acknowledgment}
%This work has been supported by National Natural Sciences Foundation of China (NSFC) under Grant 61525103 and 61771158, the Natural Scientific Research Innovation Foundation in Harbin Institute of Technology under Grant HIT. NSRIF. 2017051, and Shenzhen Basic Research Program under Grant JCYJ20160328163327348 and JCYJ20150930150304185.

\bibliography{fengbowen_2018}

\end{document}